\documentclass[final,3p,times]{elsarticle}
\usepackage{lineno,hyperref,xcolor}
\usepackage{caption}
\usepackage{booktabs}
\usepackage{cleveref}
\usepackage{multirow}
\modulolinenumbers[5]
\biboptions{numbers,sort&compress}

\hypersetup{colorlinks=true,
	linkcolor=Cerulean,%
	citecolor=Cerulean}

\journal{Journal of \LaTeX\ Templates}

\begin{document}
	
\begin{frontmatter}
	
	\title{Enhancing the long-term performance of recommender system}
	
	\author[mymainaddress]{Leyang Xue}
	
	\author[mymainaddress]{Peng Zhang\corref{mycorrespondingauthor}}
	\cortext[mycorrespondingauthor]{Correspondence and requests for materials should be addressed to P.Z}
	\ead{zhangpeng@bupt.edu.cn}

	\author[mysecondaryaddress]{An Zeng}

	\address[mymainaddress]{School of Science, Beijing University of Posts and Telecommunications, Beijing 100876, P.R. China.}
	\address[mysecondaryaddress]{School of Systems Science, Beijing Normal University, Beijing 100875, P.R. China.}
	
	\begin{abstract}
Recommender system is a critically important tool in online commercial system and provide users with personalized recommendation on items. So far, numerous recommendation algorithms have been made to further improve the recommendation performance in a single-step recommendation, while the long-term recommendation performance is neglected. In this paper, we proposed an approach called Adjustment of Recommendation List (ARL) to enhance the long-term recommendation accuracy. In order to observe the long-term accuracy, we developed an evolution model of network to simulate the interaction between the recommender system and user's behaviour. The result shows that not only long-term recommendation accuracy can be enhanced significantly but the diversity of item in online system maintains healthy. Notably, an optimal parameter $n^*$ of ARL existed in long-term recommendation, indicating that there is a trade-off between keeping diversity of item and user's preference to maximize the long-term recommendation accuracy. Finally, we confirmed that the optimal parameter $n^*$ is stable during evolving network, which reveals the robustness of ARL method.  

	\end{abstract}
	
	\begin{keyword}
		\texttt{Long-term, Recommender system, Evolution model, Robustness}
	\end{keyword}
	
\end{frontmatter}

\section{Introduction}

In the past decade, the recommender system has become an essential tool for solving the information overload problem with the rapid development of internet economy\cite{32,1}. Recommender systems provide personalized suggestions of the most relevant items for users by analyzing their historical interaction and user's profile\cite{2,3}, such as Amazon recommender\cite{4}, Tmall and so on. A number of algorithms based on various ideas and concepts have been developed to personalize the online store for each customer. These recommendation algorithms include context-based analysis\cite{14,15}, collaborative filtering\cite{5,6}, matrix decomposition\cite{7,8}, deep learning\cite{9,10} and so on. While the most of last-mentioned recommendation algorithms act on data with ratings, unary data without rating can be dealt by a class of network-based recommendation algorithms\cite{33} that represent input data with a bipartite network where users are linked with items that they have selected \cite{34}. Recently, considerable attention has been paid to network-based recommendation algorithms, such as mass diffusion (MD)\cite{11}, heat conduction(HC)\cite{12}, hybridization of mass diffusion and heat conduction\cite{13} and numerous extensions based on them\cite{16,17,18}. 

These methods perform fairly well in both accuracy and diversity in a single-step recommendation, which actually shows the short-term performance. This is one of the most important goals but not the ultimate goal for designing the recommendation algorithms\cite{19}. The original intent to use the recommender system for those internet companies is that they can yield considerable profits by making use of the niche items, because niche items enjoy higher profit margin compared to the small profit margin determined by a more competitive market of popular items\cite{35}. Hence, the ultimate goal is to broaden the scope of user's interest and recommend some niche items that are rarely purchased. Besides, user's preference changes over time as they get knowledge, maturity and experience. The online network evolves over time. Thus, a recommendation algorithm that perform very well in short-term recommendation cannot guarantee the long-term performance. Therefore, as the successive recommendation, one has to investigate the impact of algorithms on the online system and observe the long-term performance.  

Lately, some scholars begin to study the long-term performance of recommendation algorithms and find that their accuracy decreases with time if the evolution of online network fully depends on the recommendation\cite{3,19}. Many existing papers\cite{20, 21, 23} have pointed out those methods that have high accuracy in short-term recommendation tend to recommend the popular items. As online network evolves with such recommendation algorithm, most of edges are linked in small number of popular items. The extreme popular items arise and receive substantial attention in the next future, which makes the network evolve to an unhealthy state and narrows the user's choice. In this case, the long-term accuracy of recommendation algorithms that favor popular items would reduce, which is in general an intractable problem. So far, few studies has been done to futher improve the long-term performance of recommendation. Hu et al\cite{3} find that recommendation diversity is essential to keep a high long-term accuracy by increasing the length of recommendation list. Shi et al\cite{23} propose a personlized recommender based on user's preference and get better trade-off between short-term and long-term performance of recommendation. 

The major contribution of this work is to develop an evolution model of bipartite network and present a methodology called Adjustment of Recommendation List(ARL) aiming to enhance the long-term performance of recommender system. The proposed approaches are applied on the extensively analyzed dataset, enabling us to firmly compare our method with those reported algorithms. Through the empirical study, we find that the long-term recommendation accuracy can be enhanced significantly without sacrificing the short-time accuracy. Interestingly, similar to some existing literature, the diversity in long-term recommendation is also improved. By tuning the parameter incorporated into ARL, we achieve better long-term performance of recommendation algorithms. This might be an indication of there is trade-off betweeen adding diversity into evolving network and remaining original user's preference. In addition, the robustness of ARL has been confirmed. 

\section{Methods}

An online commercial system can be modeled by user-item bipartite network where users and items are represented as nodes and an edge denotes that an user has selected an item. The bipartite network can be represented by an ${M \times N}$ adjacency matrix ($M$ users and $N$ items) where the element $A_{i\alpha} = 1$ if an user $i$ has selected an item $\alpha$ and $A_{i\alpha} = 0$ otherwise. In this paper, we employed mass diffusion(MD)\cite{11} and hybridization(Hyb) \cite{13} of both mass diffusion and heat conduction as recommendation algorithms to study the long-term recommendation.

\subsection{Evolution model}

In order to observe the long-term performance, we designed an evolution model in which the evolution of online network is drived by recommendation algorithms. In the model, we assumed that user only focused on items ranked higher in recommendation list and would randomly take an item from the recommendation list of length L. In other words, users relied entirely on the recommendation when they selected some items. Thus, a new link between users and selected items was added into bipartite network. At the same time, we randomly deleted a link to keep the network size fixed. The recommendation list in next step would be generated based on the updated network with these new links. 

In the experiment, the dataset was randomly divided into training set ($E_T$) and probe set ($E_P$) according to the ratio of 90\% to 10\%. The initial network consisted of training set ($E_T$). In one macro-step of simulation, the recommendation list of each user was generated by recommendation algorithm (MD). In order to directly observe the essence of long-term recommendation accuracy as soon as possible and save computation, we set the length of recommendation list (L) as 1 so that the highest-scoring item in the recommendation list can be selected by users, namely users would take top-1 recommendation. Further, the link between the user and the selected item was made and added into the training set ($E_T$). Meanwhile, we randomly deleted a link from the user's historical record to keep the users degree fixed. This was equivalent to a so-called breaking-rewiring process. After that, the new training set ($E_{T1}$) was obtained and used to generate the recommendation list for the next macro step. We kept the probe set unchanged during the network evolution so as to enable the long-term accuracy to convergence. In each macro step, the probe set would be used to test the recommendation accuracy. Obviously, the long-term recommendation accuracy was the cumulative result of network evolution drived by the recommendation algorithms. The diagram of whole evolution process was described in \autoref{Fig.1} \emph{(a)}.

\subsection{Adjustment of Recommendation List}

The recommendation algortithms that tend to recommend the popular item for user perform well in short-term\cite{20,21,23}, e.g. mass diffusion\cite{11}, collaborative filtering\cite{5,6}. Through successive recommendations, these methods reinforce the popularity of hot items, which leads to that the system is dominated by some extremely popular items and further the long-term accuracy decreases.

\begin{figure}[!htbp]
	\centering
	\includegraphics[width=\textwidth]{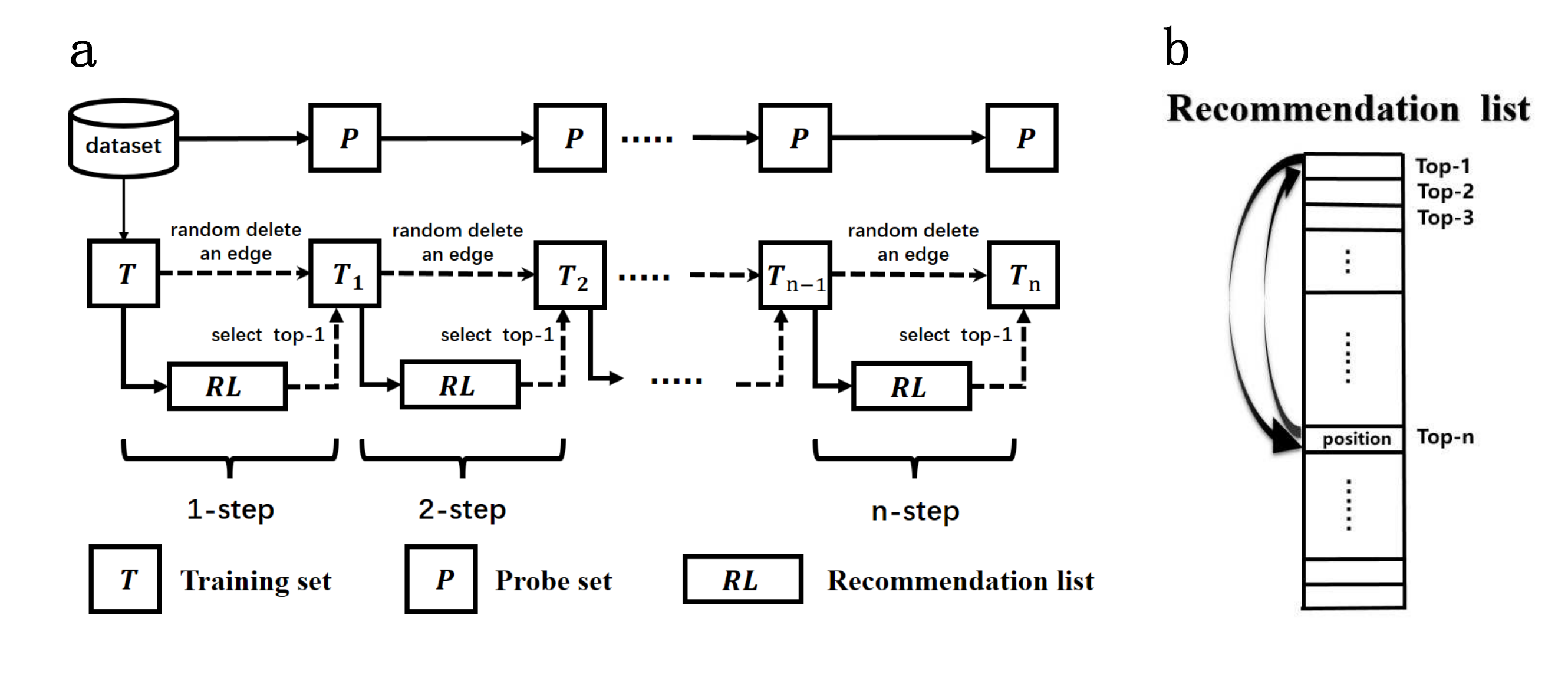}
	\caption{\textbf{\emph{(a)}} The schematic diagram of evolving network in recommendation system. \textbf{\emph{(b)}} The schematic diagram of ARL method that is implemented by exchanging the top-1 with top-n item.}
	\label{Fig.1}
\end{figure}

Therefore, we proposed a method called Adjustment of Recommendation List (ARL) to enhance the long-term recommendation performance. The main idea of ARL is to diversify personalized recommendation list to broaden the user's choose, which will transfer some of user's attention from the extremely popular items to relatively cold items. There are many ways to implement it, such as the topic diversification approach\cite{37}. For simplicity, top-1 and top-n items uncollected by users are exchanged, which forms a new recommendation list for users. A clear schematic could be seen in \autoref{Fig.1} \emph{(b)}. Here, we represented the exchange position as a parameter $n$ of ARL. A general mathematical expression could be seen in \autoref{euq}. By tuning the parameter $n$ and combining with the network evolution model, we could control the diversity introduced into evolving network to observe the long-term performance of recommendation algorithms.

\begin{equation}
\label{euq}
 RL^{'}=ARL(RL, n)                    
\end{equation}
where the $n$ denote the exchange position and $RL$ represent the recommendation list. While the $RL^{'}$ is the new recommendation list generated by ARL. $n \in [1, N_L]$, $N_L$ refers to the the number of item uncollected by users.   

\captionsetup[figure]{labelfont={bf},labelformat={default},labelsep=period,name={Fig.}}

The ARL and evolution model are essentially different. The evolution model was designed to study the adverse effect of long-term recommendation, making the recommendation accuracy significantly decrease after a number of iterative recommendations. In practice, we assumed that each user would select the top-1 item in the recommendation list and made a link between the user and the selected item. The recommendation list in next step would be generated based on the updated network with these new links. In contrast, the ARL algorithm is a method aiming for improving the recommendation. Before user making their selection, ARL algorithm re-ordered the recommendation list generated by a specific recommendation algorithm by swapping top-1 item and top-n item. This procedure would change the top-1 item in the recommendation list, further altering the item that user would select. If the recommendation list reordering is designed well, the long-term recommendation accuracy could be improved. 

\section{Experiments}

\subsection{Data}

Seven datasets were used to conducted experiment, including Movielens\cite{24} and Netflix\cite{25}, Epinions\cite{38}, Stack\cite{39}, Amazon\footnote{\url{https://www.amazon.com/}}, Delicious\footnote{\url{http://www.thedeliciousgroup.com/}} and Douban\footnote{\url{https://www.douban.com/}}. Movielens and Netflix are similar and contain the user's rating on movies. The scale of score is from 1 (i.e.worst) to 5 (i.e.best). In order to create an unweight bipartite network, we used score at least 4 as an efficient link between users and items. The movielens dataset includes 1682 users and 943 items and the resulting network contains 55375 links. The netlifx network contains 1014 users and 1977 items within 35821 links whose ratings higher than 3. Epinions is an online product rating site where users and items can be obtained. Stack Overflow is the main question and answer website of the Stack Exchange Network where nodes reprenst users and posts and an edge denotes that an user has marked a post as a favorite. Douban is a website to show the comment or remark of movies, videos and books provided by users. Amazon, Delicious and Epinions datasets is acquired by crawing website. We extracted the subsets from five original datasets (Epinions, Stack, Amazon, Delicious, Douban) to reduce the computation. A detailed information can be seen in \autoref{Table1}. In the simulation, we divided randomly the dataset into a training set $E_T$ and a probe set $E_P$. The training set contained of 90\% edges, the rest of edges constituted the probe set. Obviously, $E_T \bigcap E_P = \o{}$ and $E_T \bigcup E_P = E$. All simulation results were obtained by averaging over ten independtent experiments.

\begin{table}[!htbp]
	\setlength{\abovecaptionskip}{3pt}
	\caption{The detailed statistical characteristics of the seven experimental dataset}
	\label{Table1}
	\centering
	\small
		\begin{tabular}{ccccccc}
			\hline
			Datasets       & $N_{user}$ & $N_{item}$ & $N_{edge}$ & $<k_{user}>$ & $<k_{item}>$ & Sparsity\\ \midrule  
			Delicious      & 868   & 2,835 & 4,812   & 5.54   & 1.70  & $0.20\times10^{-2}$     \\
			Amazon         & 900   & 3,868 & 7,623   & 8.47  a & 1.97  & $0.22\times10^{-2}$     \\
			Stack Overflow & 1,199 & 1,977 & 7,886   & 6.58   & 3.99  & $0.33\times10^{-2}$     \\
			Epinions       & 1,199 & 2,978 & 48,435  & 40.40  & 16.26 & $1.36\times10^{-2}$     \\
			Netflix        & 1,014 & 1,977 & 35,821  & 35.33  & 18.12 & $1.79\times10^{-2}$     \\
			Douban         & 846   & 2,997 & 66,647  & 78.78  & 22.24 & $2.63\times10^{-2}$     \\
			Movielens      & 943   & 1,682 & 55,375 & 58.72 & 32.92 & $3.49\times10^{-2}$     \\ \bottomrule 
		\end{tabular}
\end{table}

\subsection{Metric}

\subsubsection{Ranking Score (RS)}

Ranking Score\cite{13} measures the accuracy of recommendation algorithms to generate a good ordering of items that matches the user's preference. For a target user, a recommendation list is produced by recommendation methods according user's historical record. For each item in the probe set, we can measure the rank of item in the recommendation list. A high accuracy recommendation algorithm is expected to give the item in probe set a higher rank, which leads to a small ranking score. Ranking score of a target user is obtained by averaging over all entries in the probe set to quantify the recommendation accuracy of method. A formula of ranking score is as follow:
\begin{equation}
RS_{u\alpha} =\frac{l_{u\alpha}}{L_{u}},
\end{equation}
where $l_{u\alpha}$ is the rank of item $\alpha$ in the recommendation list of user $u$. $L_{u}$ denotes the number of uncollected items, namely the length of recommendation list in offline testing. The ranking score of the whole system is obtained by averageing $RS_{u\alpha}$ over all users. Obviously, $RS \in (0,1)$. The smaller the ranking score, the higher the recommendation accuracy of algorithms.  In this work, there is a case where user-item pairs in the probe set appear in the evolving network, resource score of items equal to zero. In order to calculate reasonably the ranking score, these items are put into the recommendation list and are ranked last, which can be regarded as those items selected cannot be recommended in current macro step. In fact, these user-item pairs may be removed from evolving network in next macro step. In actual calculation, the formula of ranking score as follow: $RS_{u}=\sum_{\alpha\in probe_{u}}\frac{{l}'_{u\alpha}}{L}$, where ${l}'_{u\alpha}$ is the rank of probe-set item $\alpha$ in the recommendation list, $L$ means the number of all items in system. The value obtained from $RS_u$ underestimated original RS, but it doesn't matter for the comparison of long-term accuracy between mass diffusion and ARL.

\subsubsection{Gini coefficient}

We exployed the Gini index\cite{26} to measure the heterogeneity of item popularity distribution in long-term recommendation. The Gini index is originally proposed to assess income inequalities of inhabitants or families in a country\cite{30}, it has been widely used to measure the dispersion in other fields\cite{27,28,29,36}. The following equation can be applied to calculate the Gini index:
\begin{equation}
G = \frac{2\sum_{\alpha=1}^{N}{\alpha k_{\alpha}}}{N\sum_{\alpha=1}^{N}{k_{\alpha}}} - \frac{N+1}{N},
\label{equation}
\end{equation}
where $k_{\alpha}$ denotes the degree of item $\alpha$, representing the popularity of item $\alpha$ in the system. The $\alpha = 1$ to $N$, that are indexed in non-decreasing order ($k_{\alpha} \leq k_{\alpha+1}$). The $N$ is the number of items. Hence, the value of Gini index ranges between zero and one , which corresponds to the equal popularity of item (i.e. every item has the same degree) and completely unequal popularity of item (i.e. only an item is selected by users, while items else have zero degree) respectively. A higher Gini cofficient indicates a more heterogeneous distribution, and vice versa. In some sense, the Gini index can be regarded as an health indicator of the whole system. For example, all edges are linked to one item and other items have no links when Gini index equals to one. In this extreme case, the diversity of item in the system is very poor and there are no valid and valuable information that be used to make recommendations. In addition, the heterogeneity of item popularity distribution reflects the diversity of item in system. 

\subsubsection{Jaccard index}

Jaccard index\cite{31} is proposed to measure the similarity between finite sample sets more than one hundred years ago. Here, we use Jaccard index to calculate the similarity between top-1 and top-n item. The formula of Jacard index is as follow: 
\begin{equation}
	S_{\alpha\beta}^{Jaccard}= \frac{\left | \Gamma(\alpha)\cap \Gamma(\beta) \right |}{\left|\Gamma(\alpha)\cup\Gamma(\beta)\right|}. 
\end{equation}
For each item $\alpha$, $\Gamma(\alpha)$ denotes the set of neighboor of item $\alpha$, namely the set of user that have selected the item $\alpha$. $|\Gamma(\alpha)\cap\Gamma(\beta)|$ represent the common neighoors between item $\alpha$ and $\beta$. The value of Jaccard index ranges from 0 to 1.

\begin{figure}[!htbp]
	\centering
	\includegraphics[width=\textwidth]{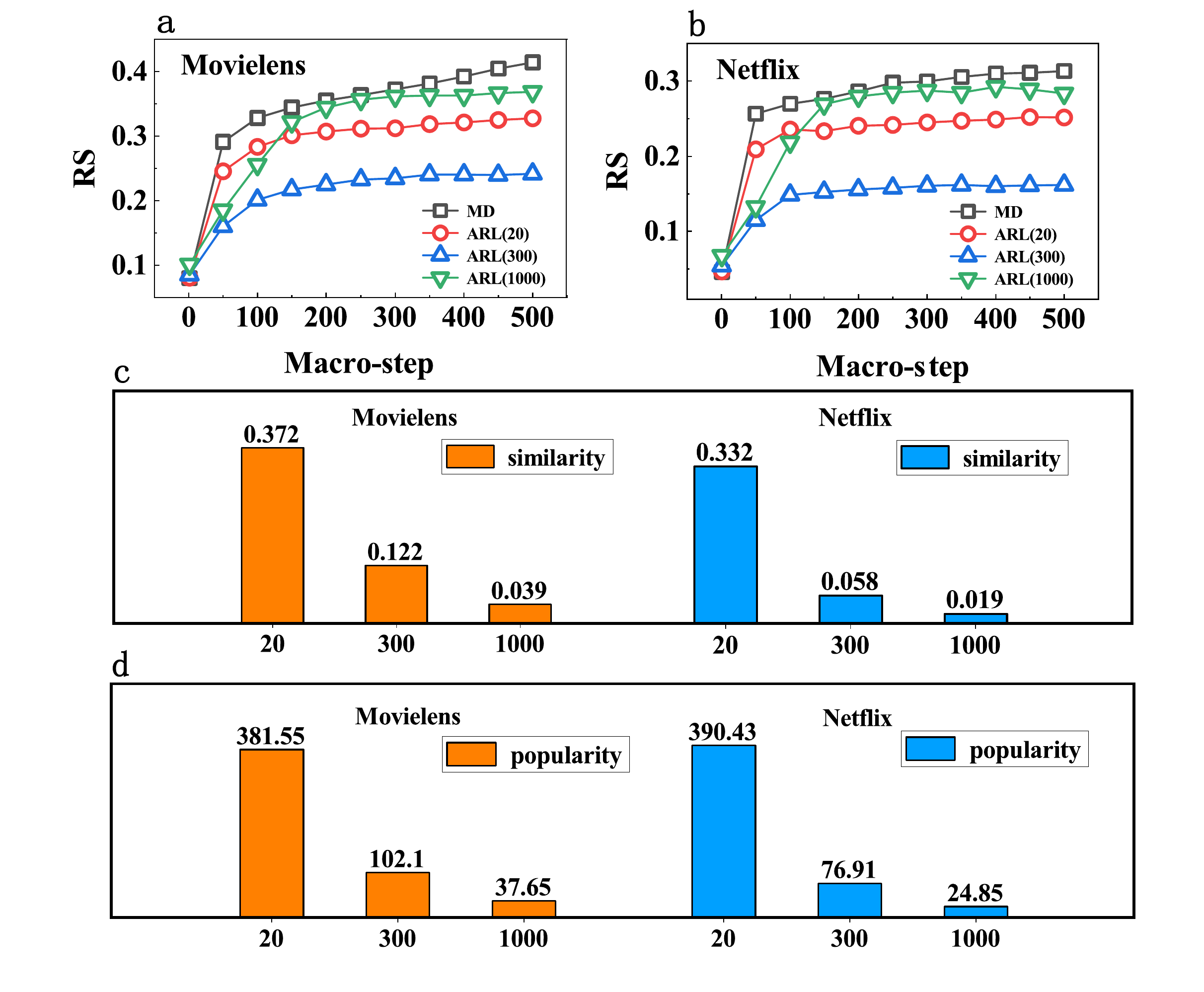}
	\caption{\textbf{\emph{(a)(b)}} The ranking score plotted as a function of macro-step for different parameter $n$ of ARL on Movielens and Netflix data sets, showing the prediction accuracy under different period of recommendation. Here, n = 1 degenerates to original mass diffusion. \textbf{\emph{(c)}} The similarity of item between the top-1 and top-$n$, which is measured by Jaccard index \cite{31}. The horizontal axis represents the parameter $n$ of ARL, namely exchange position.  \textbf{\emph{(d)}} The popularity of top-$n$ item. The popularity is calculated by degree of item. The greater the popularity of the item, the smaller the diversity it introduces into the evolving network. The similarity and popularity of item are averaged over macro-step of network evolution.}
	\label{Fig.2}	
\end{figure}
   
\section{Results}

In this section, we conducted a set of experiments to examine the long-term recommendation performance of ARL. The ranking score and Gini index was used to measure the recommendation accuracy and the heterogeneity of item popularity distribution in the system respectively. We employed the Jaccard index to evaluate the similarity between top-1 and top-$n$ item. 

\subsection{The long-term recommendation performance of ARL}

The parameter $n$ has a significant impact for the long-term recommendation performance. For instance, the similarity between the top-1 and top-n items will be lower if the value of $n$ is very large, further leading to introducing more diversity into the evolving network. At the same time, this also results in loss of user's preference, which still reduces the long-term recommendation accuracy. Therefore, it is important for long-term recommendation to choose the appropriate parameter.

\begin{figure}[!htbp]
	\centering
	\includegraphics[width=\textwidth]{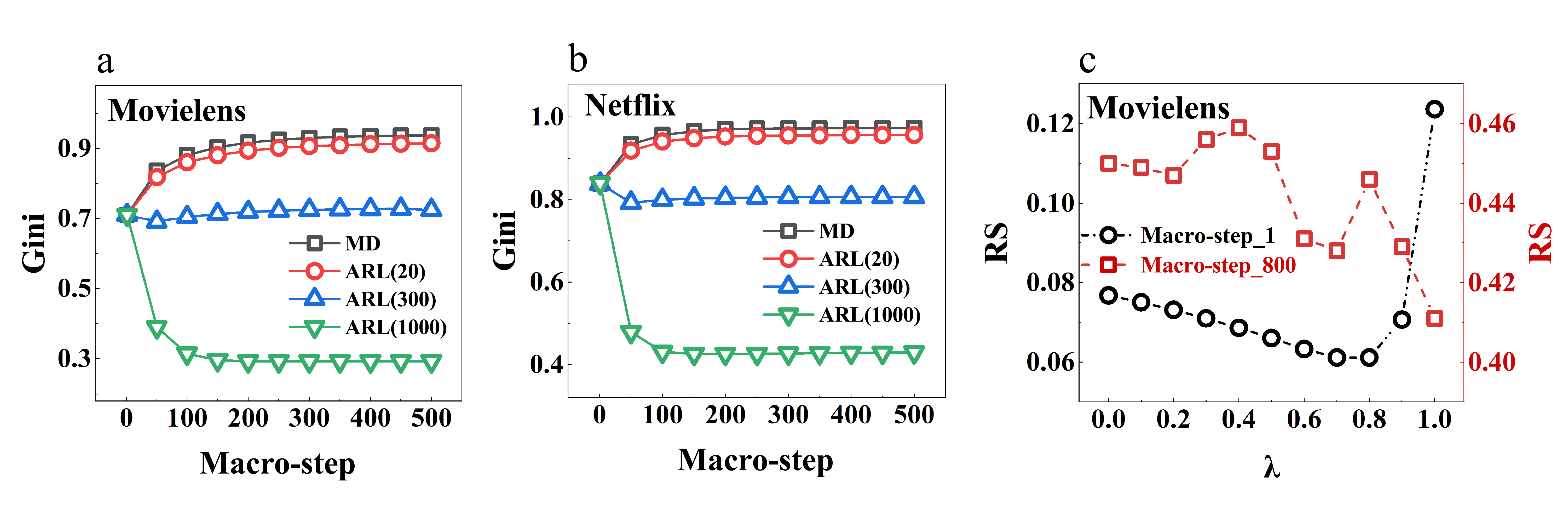}
	\caption{\textbf{\emph{(a)(b)}} The Gini index plotted as function of macro-step for different parameter $n$ on Movielens and Netflix.  \textbf{\emph{(c)}} The ranking score plotted as function of $\lambda$ for hybrid algorithm under 1-step and  800-step. $\lambda = 0$ where the hybrid method equals to mass diffusion, $\lambda = 1$ refers to the pure heat conduction algorithm. }
	\label{Fig.3}
\end{figure}

We showed the recommendation accuracy of ARL conducted on Movielens and Netflix for different n under different macro-step in \autoref{Fig.2}\emph{(a)(b)}. Note that the ARL(1) degenerates to the original mass diffusion. We found that the curve of four algorithms tended to reach a stable after 250 macro-step in the \autoref{Fig.2}\emph{(a)(b)}. Therefore, we regarded the average of RS from 250 to 500 macro-step as long-term recommendation accuracy. In the \autoref{Fig.2}\emph{(a)}, one could see that ARL(300) has lowest accuracy loss in long-term recommendation. Compared with the mass diffusion, ARL(300) has improved the long-term recommendation accuracy of 38\%, which confirmes that ARL method we proposed is effective (\emph{More results could be seen Fig.6-10 (a), Supplementary}). However, the improvement of long-term accuracy for ARL(20) and ARL(1000) are not obvious although they can enhance the long-term recommendation accuracy. A possible explaination can be seen in \autoref{Fig.2}\emph{(c)(d)}. As expected, the similarity of items between the top-1 and top-20 are relative high, more relevant items can be recommended for users with the network evolution. Meanwhile, the top-20 item is also very popular, indicating that less diversity is introduced into the evolving netwotk. The joint effect of two factors determine that most links in the resultant network are still connected to the popular items (this can also be confirmed in \autoref{Fig.3}\emph{(a)}), leading to the lower improvement of long-term recommendation accuracy. While the similarity between the top-1 and top-1000 item and popularity of top-1000 item are extremely low, which reveals that too many irrelevant niche items are involved in evolving network and make the resultant network contain less valuable information about user's real preference. This might explain that the enhancement of ARL(1000) in the long-term recommendation is not obvious. Therefore, it is essential for keeping a high long-term recommendation accuracy to both keep a certain diversity and remain user's preference. The same result can also be proved in \autoref{Fig.2}\emph{(b)}.

We studied the health state of the system in the long-term recommendation. The Gini index was employed to measure the heterogeneity of item popularity distribution during the network evolution. We showed the Gini index of ARL for different parameters under different macro-step in \autoref{Fig.3}\emph{(a)(b)}. The Gini index of MD and ARL(20) increase gradually with the macro-step and finally reache stable, suggesting the item popularity become more heterogeneous with successive recommendation. Moreover, the $ARL(20)$ still result in the rise of extremely popular item and reduce item diversity in long-term recommendation. However, the Gini index of $ARL(1000)$ dramatically decreases with the macro-step and eventually reaches values lower than those produced by other algotithms, which indicate that more links are connected to the niche items with network evolution. Interestingly, the Gini index of $ARL(300)$ keeps the same as the short-term. This might be an important indication to matain the higher accuracy in long-term recommendation. A reasonable explanation supported in \autoref{Fig.2}\emph{(c)(d)} is that there is a trade-off between introduction of diverse items and retention of user's perference to enhance the long-term recommendation performance. The diversity of item in different online systems could be improved by ARL (\emph{Fig.6-10 (b), Supplementary})  

\begin{figure}[!h]
	\centering
	\includegraphics[width=\textwidth]{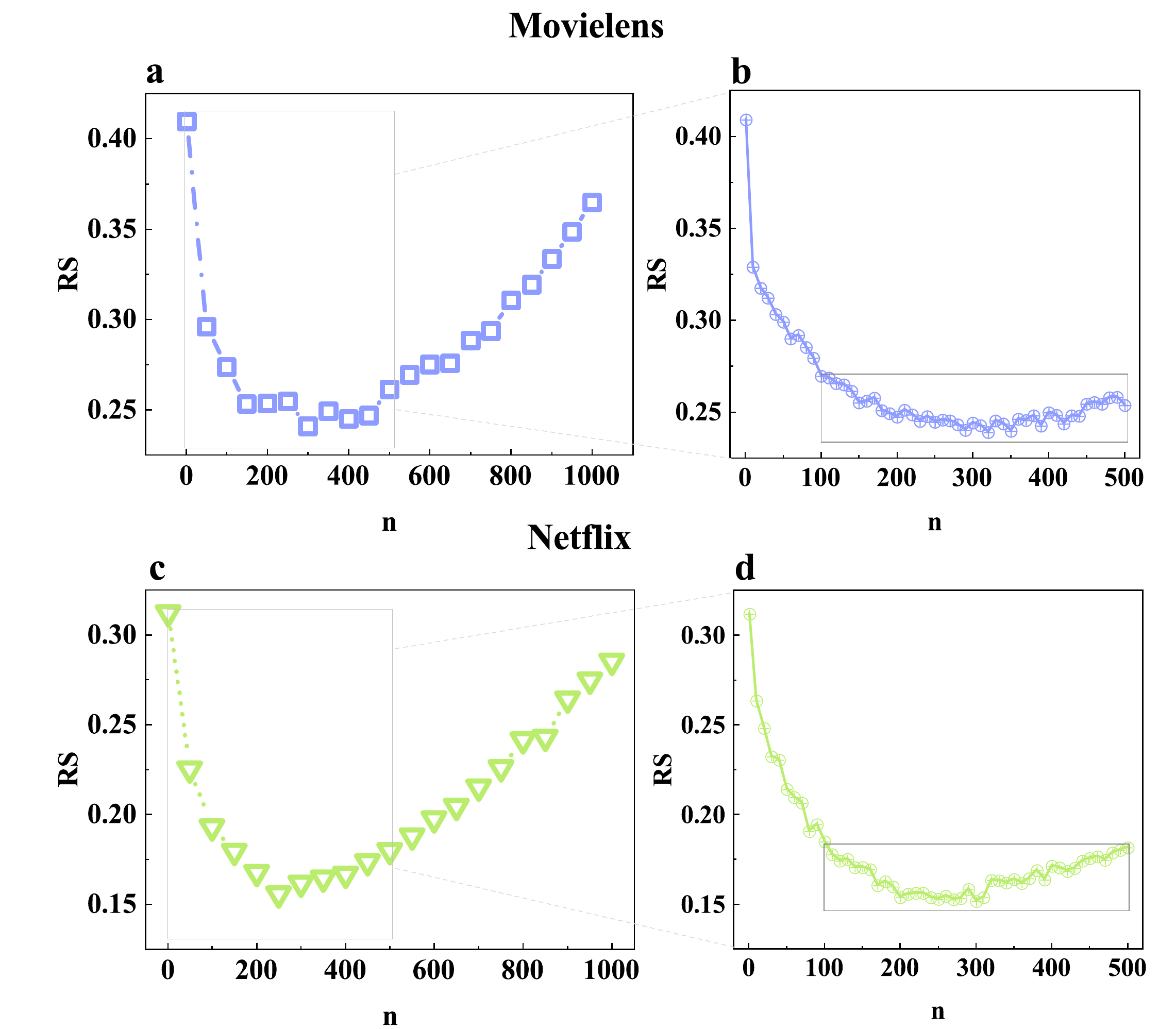}
	\caption{The ranking score (RS) plotted as a function of $n$. The RS value of point is calculated by averaging over RS from the 450 to 500 macro-step, showing the long-term recommendation accuracy of ARL($n$). \textbf{\emph{(a)(c)}} The long-term recommendation accuracy is shown on Movielens and Netflix with a resolution of 50, which enable us to capture the trend in parameter space. \textbf{\emph{(b)(d)}} The long-term recommendation accuracy is shown on Movielens and Netflix with a resolution of 10. One could see the difference of long-term accuracy is fairly small for those points contained by rectangle. }
	\label{Fig.4}
\end{figure}

\subsection{The long-term recommendation accuracy of hybrid recommendation method}

The hybrid method of mass diffution and heat conduction\cite{15} was employed to observe the long-term recommendation accuracy. Through tuning the parameter $\lambda$, the optimal recommendation accuracy could be obtained. We mainly focused on how the optimal parameter ($\lambda^{*}$) changes in long-term recommendation, showing the RS of both 1 step and 800 steps in \autoref{Fig.3}\emph{(c)} which corresponded to the short-term and long-term recommendation accuracy respectively. The $\lambda^{*}$ is 0.8 when the short-term recommendation accuracy achieve the maximum, whereas the $\lambda^{*}$ is 1 in long-term recommendation. With the network evolution, the $\lambda^{*}$ shifts to a larger value corresponding to the heat conduction method. This is a natural result because the heat conduction tends to recommend the niche item for users. The analysed result again confirms that the diversity of item is essential to enhance the long-term recommendation accuracy.   

\subsection{The optimal parameter of ARL}

In order to maximize the long-term recommendation accuracy, we performed the experiment for different parameters $n$ with a step of 50 from 0 to 1000. The $n$ = 1 denotes that the result equals to the long-term recommendation accuracy of mass diffusion. For different parameters $n$, the recommendation accuracy reachs stable after 450 steps. Therefore, we regarded the average of accuracy from 450 to 500 steps as the long-term recommendation accuracy.  The scatter plot of long-term recommendation accuracy of ARL as a function of $n$ could be seen in \autoref{Fig.4}. We found that the value of RS decreased sharply with $n$ and reach the lowest at 320, then it slowly increased in \autoref{Fig.4} \emph{(a)}. This suggested that ARL(320) can maximize the long-term recommendation accuracy according to current resolution of $n$, which can be interpreted as the trade-off between the introduction of item diversity and retention of user's preference information. At the optimal $n^*$, the long-term recommendation accuracy has been enhanced siginificantly compared with the mass diffusion algorithm. Except for the $n^{*}$, others parameter around the $n^{*}$ still maintains relatively higher accuracy (see those points contained by rectangle in \autoref{Fig.4}\emph{(b)}), confirming that the ARL method we proposed is very robust. The same conclusion can be obtained in \autoref{Fig.4} \emph{(b)}. The optimal parameter of ARL is also analyzed on other datasets (\emph{Fig.6-10 (c), Supplementary}).

\begin{figure}[!h]
	\centering
	\includegraphics[width=\textwidth]{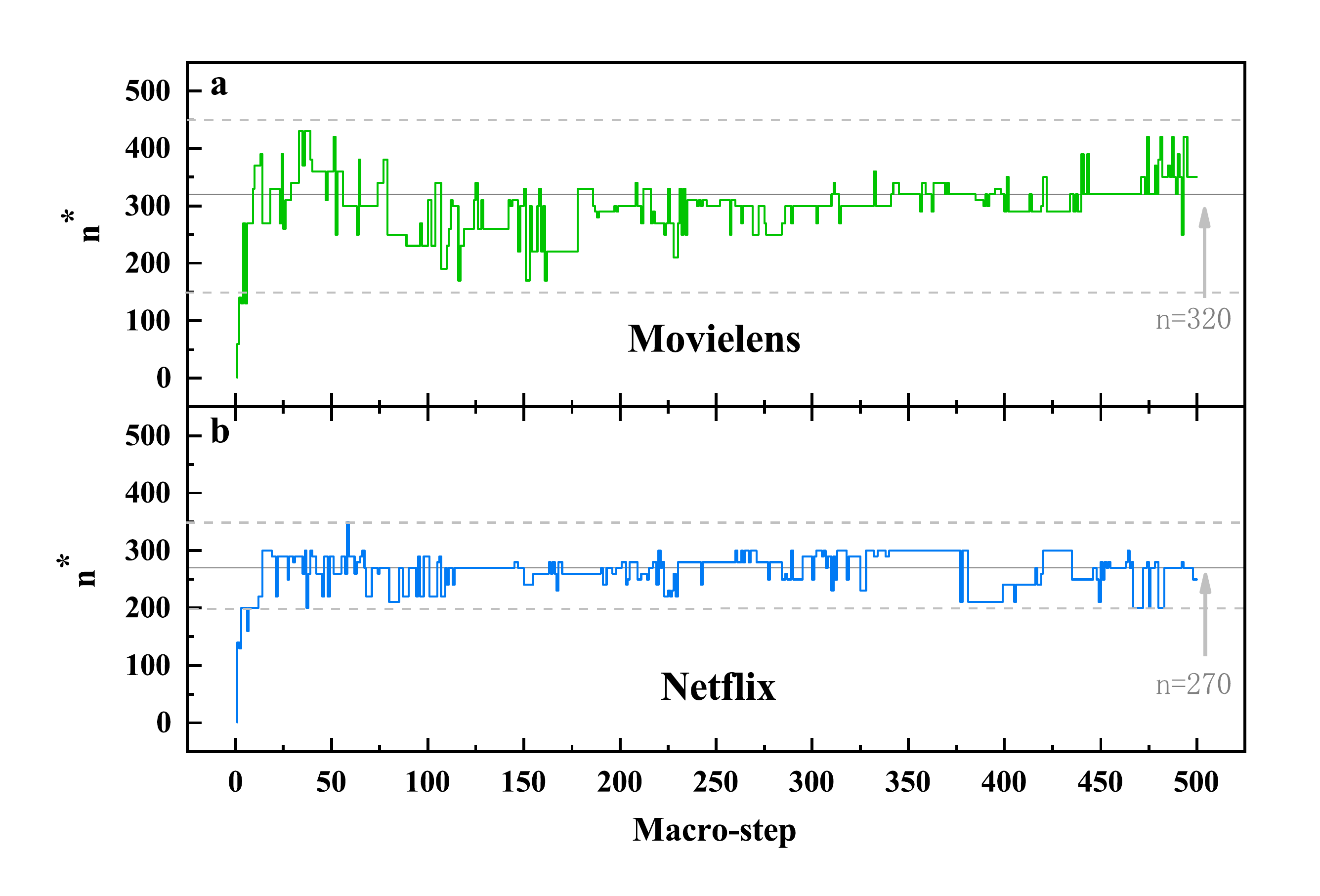}
	\caption{The step plot of optimal parameter ($n^{*}$) in different maro steps, which shows the stability of parameter for different periods of recommendation. Note that the ARL($n^{*}$) refers to the highest recommendation accuracy in current macro step. The gray line shown in the figure is to guide eyes where most $n^{*}$ in different macro steps appear at this line or parameters fluctuate around the gray line. The dotted line represents the boundary of parameters in the long-term recommendation. \textbf{\emph{(a)}} Movielens \textbf{\emph{(b)}} Netflix}
	\label{Fig.5}
\end{figure}

We finally showed the optimal parameter $n^{*}$ to maximize the accuracy in different macro steps. As is shownin the \autoref{Fig.5}\emph{(a)(b)}, the parameter $h^{*}$ gradually rises with network evolution. Then, it tends to fluctuate around the gray line that respresents the value of optimal parameter ($n^{*}$) in long-term recommendation. The result suggests that less diversity should be introducted into the evolving network to maximize the accuracy in the short-term recommendation. With the increasement of macro step, more diversity should be added into the evolving network. Besides, the optimal parameter $n^{*}$ keeps within a certain range for different macro steps (see the two dotted line in \autoref{Fig.5}\emph{(a)(b)}), which make the ARL maintain the higher recommendation accuracy for different macro steps by fixing a parameter, For example, the optimal parameter $n^{*}$ under different macro steps is around 270 on Netflix dataset. Thus, ARL(270) can achieve more higher recommendation accuracy in whole evolving network. This indicate that the optimal parameter of ARL has strong stability. These results have been confirmed on different datasets (\emph{Fig.6-10 (d), Supplementary}). As the ARL is applied on different data sets, we find the $n^*$ is related to the sparsity of data set, namely the denser the data set, the greater the value of $n$ (\emph{Table.2 and 3, Supplementary}).

\section{Conclusion}

Numerous recommendation methods have been developed to extend the personalized selection. Although some attention has been paid to the long-term recommendation performance, there are still much unknown for the improvement of the long-term recommendation performance. In this paper, according to the assumption that users only keep their eyes on those items ranked higher in recommendation list, we proposed an evolution model to simulate successive recommendations between the recommender system and users, further observing the long-term performance of recommendation algorithms. We found the prediction accuracy of mass diffusion gradually decreased with the network evolution when users taken top-1 recommendation. Thus, the ARL method was proposed to enhance the long-term accuracy of recommender system. The top-1 and top-n item are exchanged to diverse the recommendation list, which make the cold item be introduced into the evolving network. Although some cold items are selected, it is usually considered that the accuracy might be decreased. In fact, cold items selected also enrich the diversity of item in online system, which enable the initial resources of network-based diffusion algorithms to cover more items and enlarge the range of potential recommended item. Many existing papers\cite{13} has also pointed out that one can improve both in accuracy and diversity with a well-designed recommendation algorithm. This is because the higher accuracy is resulted from precisely predicting cold items liked by users. Interestingly, similar to some existing literature, we found both the long-term accuracy and diversity in online system was improved by ARL. Besides, by tuning the parameter $n$, we showed there was a trade-off between the introduction of item diversity and remaining of user preference to maximize the long-term recommendation accuracy. Meanwhile, the optimal parameter $n^*$ fluctuated within a certain range in different period of recommendation, which demonstrates the ARL method is very robust and stable.

The enhancement of long-term recommendation performance can not only satisfy the personalized need of the people but increase the pofit of commercial system. Here, we proposed a novel frame ARL to achieve higher long-term recommendation accuracy, which can be widely applied to those recommendation algorithms that tend to recommend the popular items by tuning the parameter $n$. Besides, there are a number of interesting extensions that could be done in the future. On one hand, a lot of realistic factors would be considered into the evolution model, such as the change of user preference with time (updating user preference based on user's online click rate),  the extent to which users rely on the recommendation, and the influence of social relationships among the users and so on. On the other hand,  we could design new recommendation algorithms combining the short term and long term. For example, the mechanism of user preference recession over time can be introduced into the recommendation algorithm. 

\section*{Acknowledgements}
This work is supported by the National Natural Science Foundation of China [Grant No.61403037, No.61603046], the Natural Science Foundation of Beijing [Grant No.L160008].

\bibliography{bibfile}
	
\section*{\LARGE \centering{Supplementary Material \\Enhancing the long-term performance of recommender system}}
\centerline{\large{Leyang Xue, Peng Zhang, An Zeng}}
\hfill

In this paper, we apply the ARL method on different datasets and compare the long-term recommendation accuracy with original mass diffusion. In the Delicious, Amazon, Stack Overflow, Epinions and Douban date sets, we show the ranking score, Gini index and the optimal parameter $n^{*}$ of ARL under different macro steps as well as the long-term recommendation accuracy of ARL(n). One can see that long-term performance of mass diffusion can be enhanced by ARL. Then, the effectives and robustness of ARL have been confirmed. Besides, we analyze the impact of different datasets on $n^{*}$. 
	
\captionsetup[figure]{margin=10pt, font= small, format = plain, labelfont={bf},labelformat={simple},labelsep=endash,name={Fig.}}

\begin{figure}[ht]
	\centering
	\includegraphics[width=\textwidth]{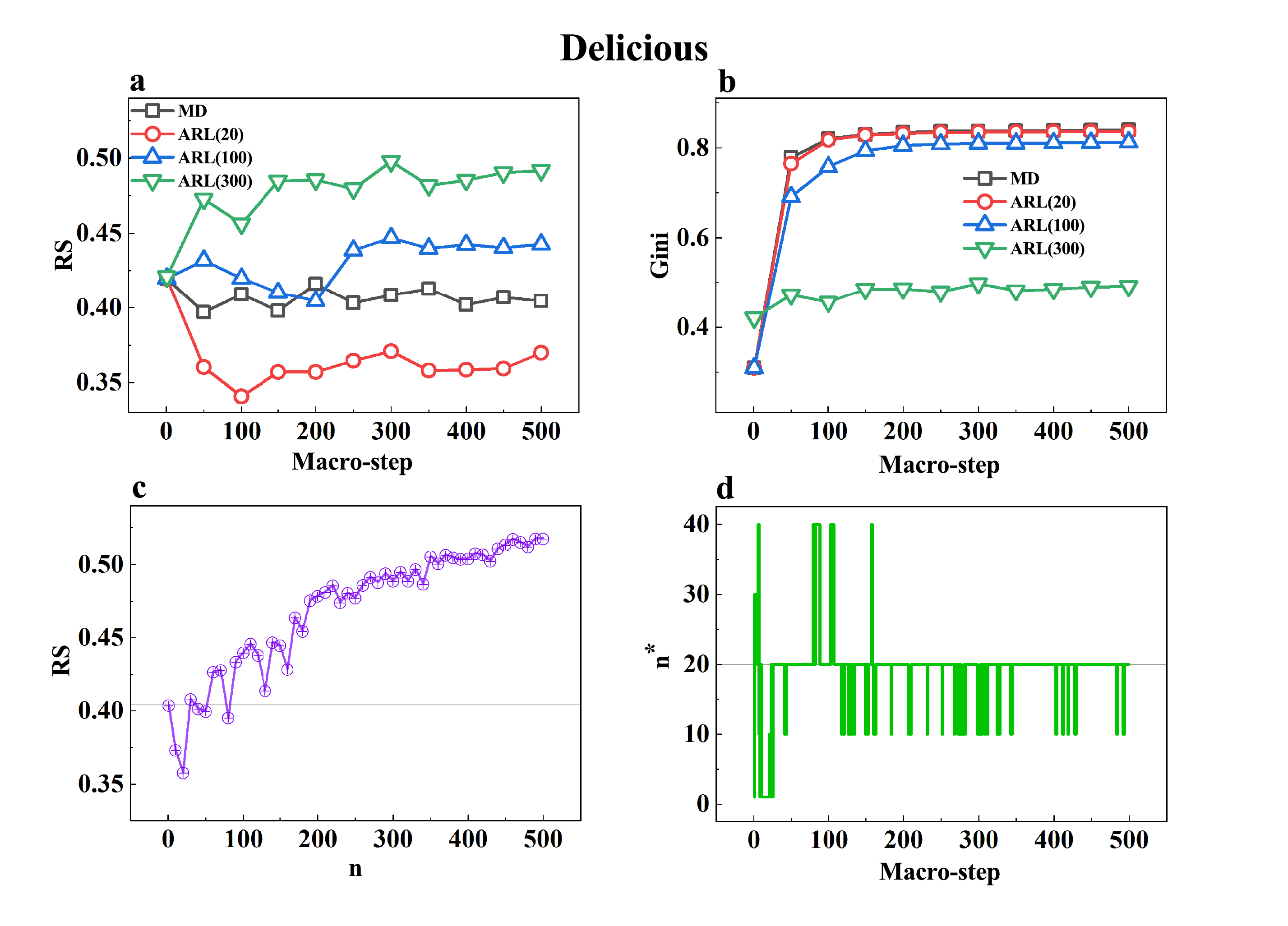}
	\caption{Delicious. \textbf{\emph{(a)}} Ranking Score(RS) plotted as a function of macro-step for ARL($n$), showing the prediction accuracy under different period of recommendation. Note that the ARL(1) degenerates to the mass diffusion (MD). \textbf{\emph{(b)}} Gini coefficient plotted as a function of macro-step for ARL($n$). \textbf{\emph{(c)}} Ranking score(RS) plotted as a function of $n$ with a resolution of 10, where the point of RS is calculated by averaging over RS form 450 to 500 macro-step. The purple curve and gray line respectively represent the long-term recommendation accuracy of both ARL($n$) and original mass diffusion. \textbf{\emph{(d)}} The step plot of optimal parameter ($n^{*}$) in different macro steps. Note that the ARL($n^{*}$) refers to the highest recommendation accuracy in current macro step. The gray line shown in the figure is to guide eyes where most $n^{*}$ in different macro-steps appear at this line or parameters fluctuate around the gray line.}
	\label{Fig.S1}
\end{figure}

\section*{The analysis of ARL on different datasets}
		
\subsection*{\centering{Delicious}}	
In Delicious data set, the accuracy for n=20,100,300 under different macro steps can be observed in \autoref{Fig.S1} (a). Interestingly, the value of RS of mass diffusion in long-term recommendation sightly lower than single-step recommendation, which suggests that long-term accuracy is improved compared with short time. The result is mainly caused by extreme sparse data set. This is because the coverage of initial resource of mass diffusion is limited in sparse network\cite{40}. By the evolution model and ARL, diverse items are added into evolving network and the link is rewired, which enable resource to cover more items and further could predict items in the probe set. Even though in such a case, the long-term recommendation accuracy still can be enhanced by ARL(20). The same result can be seen in \autoref{Fig.S1} (c), there are some parameters in long-term recommendation whose accuracy higher than mass diffusion. In the \autoref{Fig.S1} (d), most optimal parameters of ARL appear at 20, especially for the long-term recommendation (from 400 to 500 macro-steps), which confirms the parameter stability of ARL under different period of recommendation .
	
\begin{figure}[!h]
	\centering
	\includegraphics[width=\textwidth]{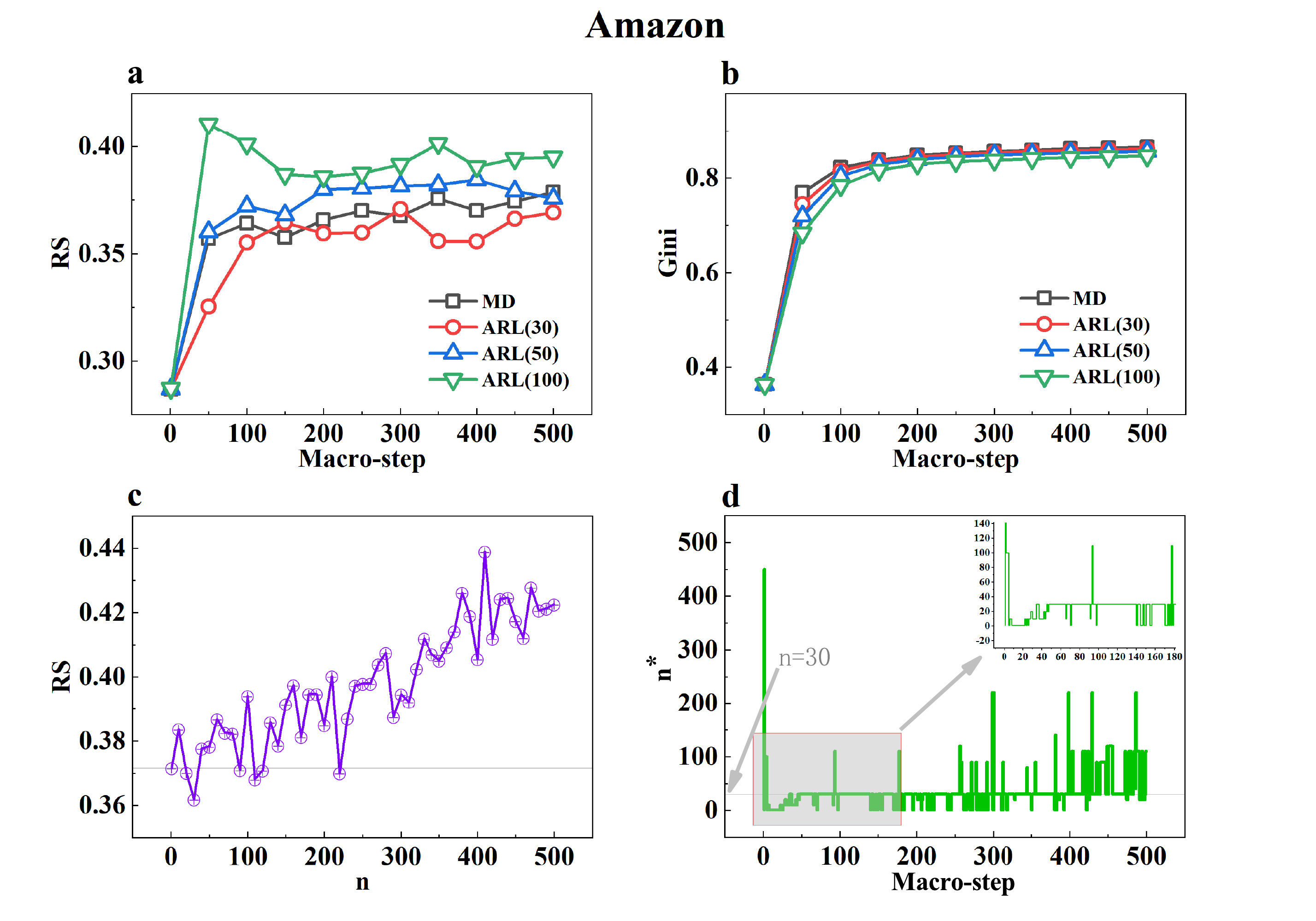}
	\caption{Amazon.  \textbf{\emph{(a)(b)(c)(d)}} The description is the same as the caption of Fig.6 .}
	\label{Fig.S2}
\end{figure}

\subsection*{\centering{Amazon}} 
		
In \autoref{Fig.S2}, we show the result of ARL conducted on Amazon data set. In \autoref{Fig.S2} (c), we find that there are some parameters appearing below the straight line whose long-term accuracy higher than original mass diffusion. The result suggests the ARL method is effective to improve the long-term accuracy. According the indication that long-term recommendation accuracy achieves the highest at n = 30, we set the n as 30,50,100 and plot the RS and Gini as a function of different macro-step, which is shown in \autoref{Fig.S2} (a) and (b) respectively. The enhancement of long-term performance for n=30 is not obvious in Amazon compared with other datasets, especially for the Gini. This is because most users selected less items and a little number of users bought more niche items in sparse dataset\cite{40}. In such condition, the improvement of Gini is limited by breaking-rewiring process of link. From the \autoref{Fig.S2} (d), one can see the most optimal parameters appear at n = 30 in different macro steps, which reveals ARL(n) is very robust rather than only perform well at specific recommendation period. 

\begin{figure}[!ht]
	\centering
	\includegraphics[width=\textwidth]{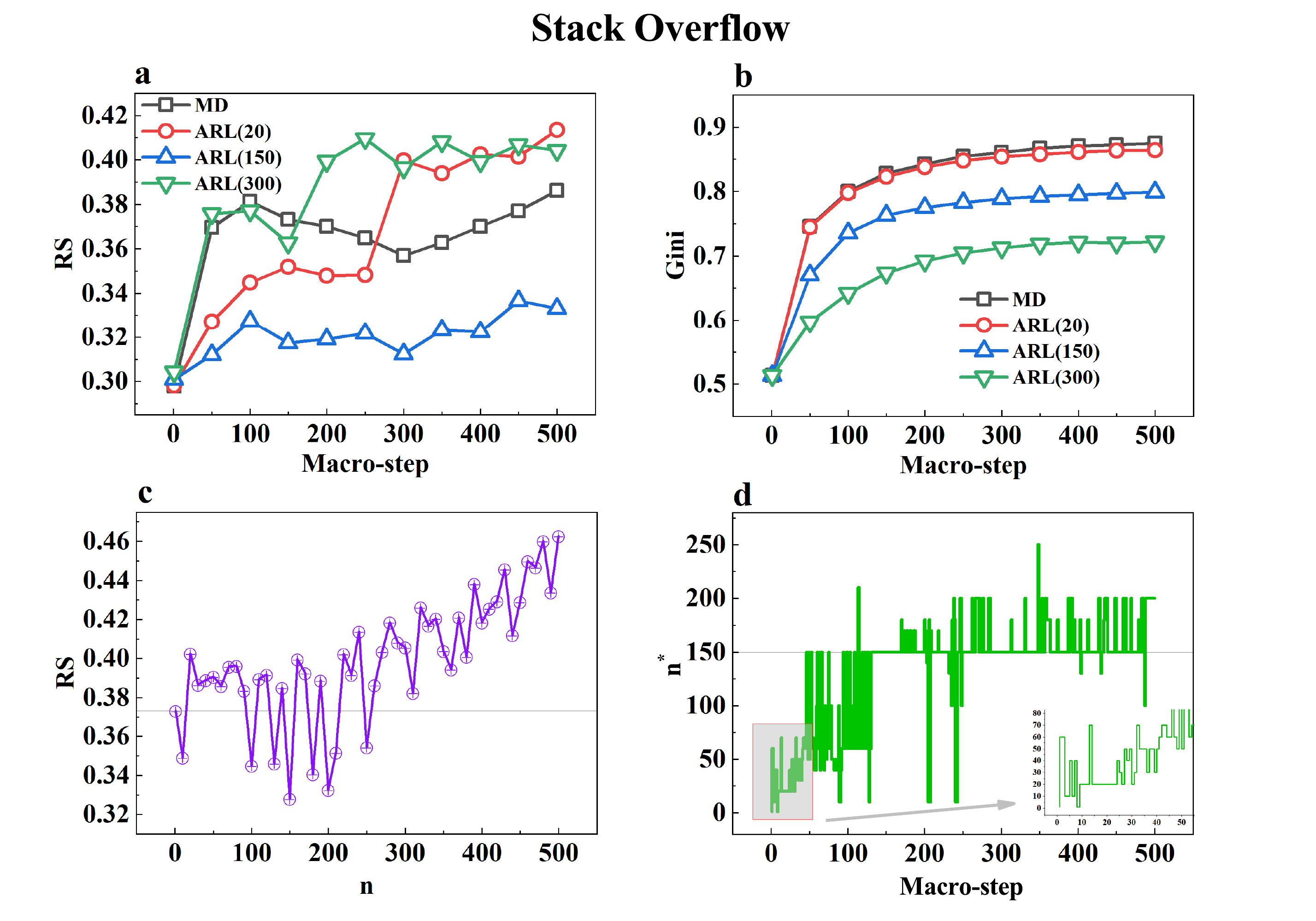}
	\caption{Stack Overflow. \textbf{\emph{(a)(b)(c)(d)}} The description is the same as the caption of Fig.6 .}
	\label{Fig.S3}
\end{figure}
	
\subsection*{\centering{Stack Overflow}}

The data of Stack Overflow is more dense than Delicious and Amazon. In the \autoref{Fig.S3} (c), the highest accuracy in long-term recommendation appear at n = 150. Compared with the optimal parameter in sparse data, the value of n gradually increase, this is a indication of more diverse item need to be introduced into the evolving network. Besides, the improvement of accuracy and Gini is obvious under different macro steps by ARL(150) observed in \autoref{Fig.S3} (a)(b). From the \autoref{Fig.S4} (d), we find the value of $n^{*}$ continually increases from 1 to 50 macro step (As is shown in the inset) and fluctuate around the line of n=150 after 150 macro step. This suggest that we need to put the less diverse items into evolving network in short-term recommendation and more diversity in long-term recommendation. 
	
\subsection*{\centering{Epinions}}
	
\begin{figure}[!ht]
		\centering
		\includegraphics[width=\textwidth]{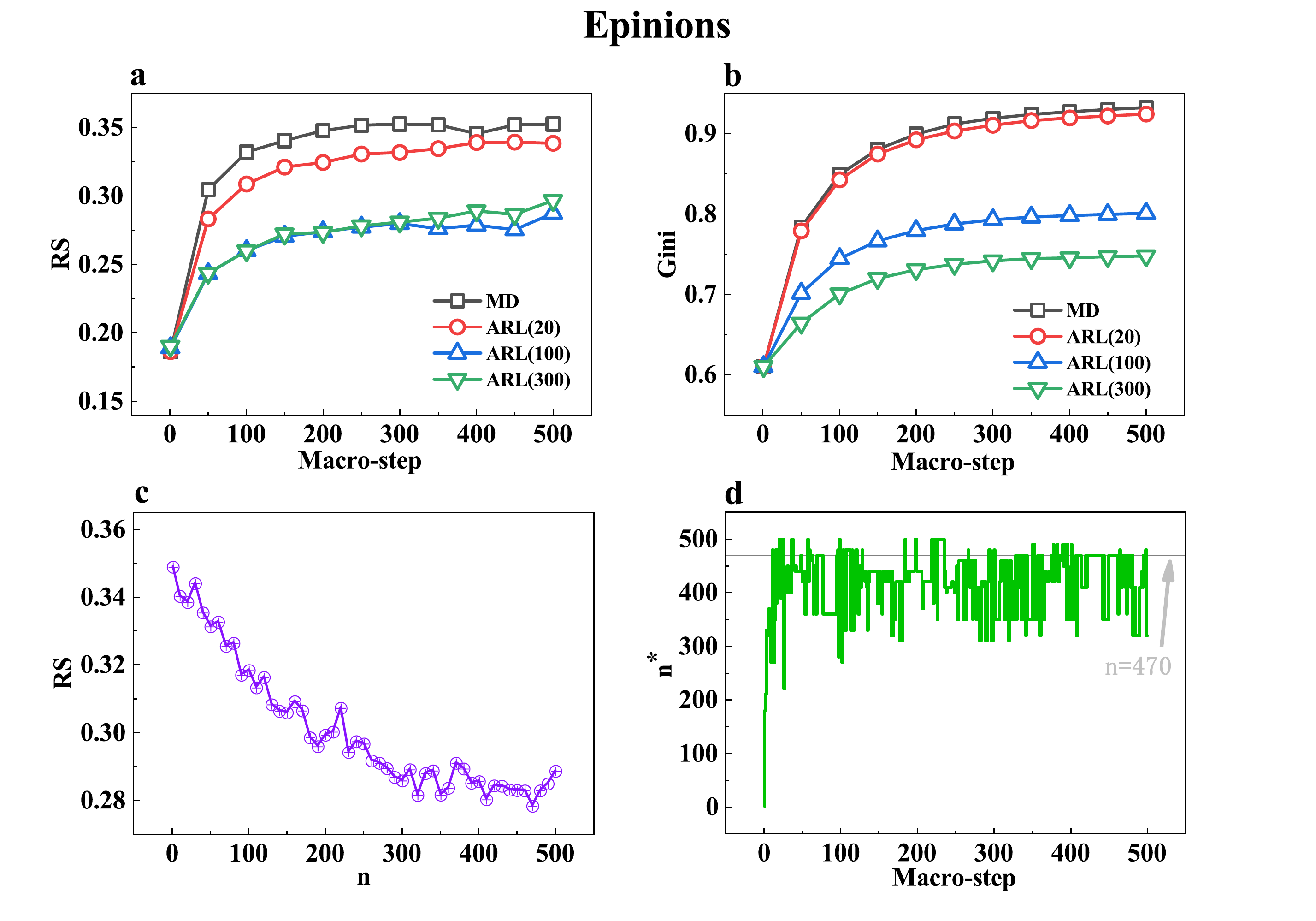}
		\caption{Epinions. \textbf{\emph{(a)(b)(c)(d)}} The description is the same as the caption of Fig.6 .}
		\label{Fig.S4}
\end{figure}
	
The Epinions data is more dense than previous analysed data. The ARL performs very well in Epinions dataset. As is shown in \autoref{Fig.S4} (c), the accuracy of all parameters in long-term recommendation higher than mass diffusion. Moreover, the lowest value of the curve appear at n= 470. One can see the the long-term accuracy can be enhanced significantly by ARL(470). The similar result can be seen in \autoref{Fig.S4} (a)(b). These results show our method is effective for improving the accuracy and diversity under different macro steps. In the \autoref{Fig.S4} (d), the difference of long-term accuracy between 300 and 500 is rather slight although the $n^{*}$ in different macro steps fluctuate around n=400.
	
\subsection*{\centering{Douban}}
	
\begin{figure}[!ht]
		\centering
		\includegraphics[width=\textwidth]{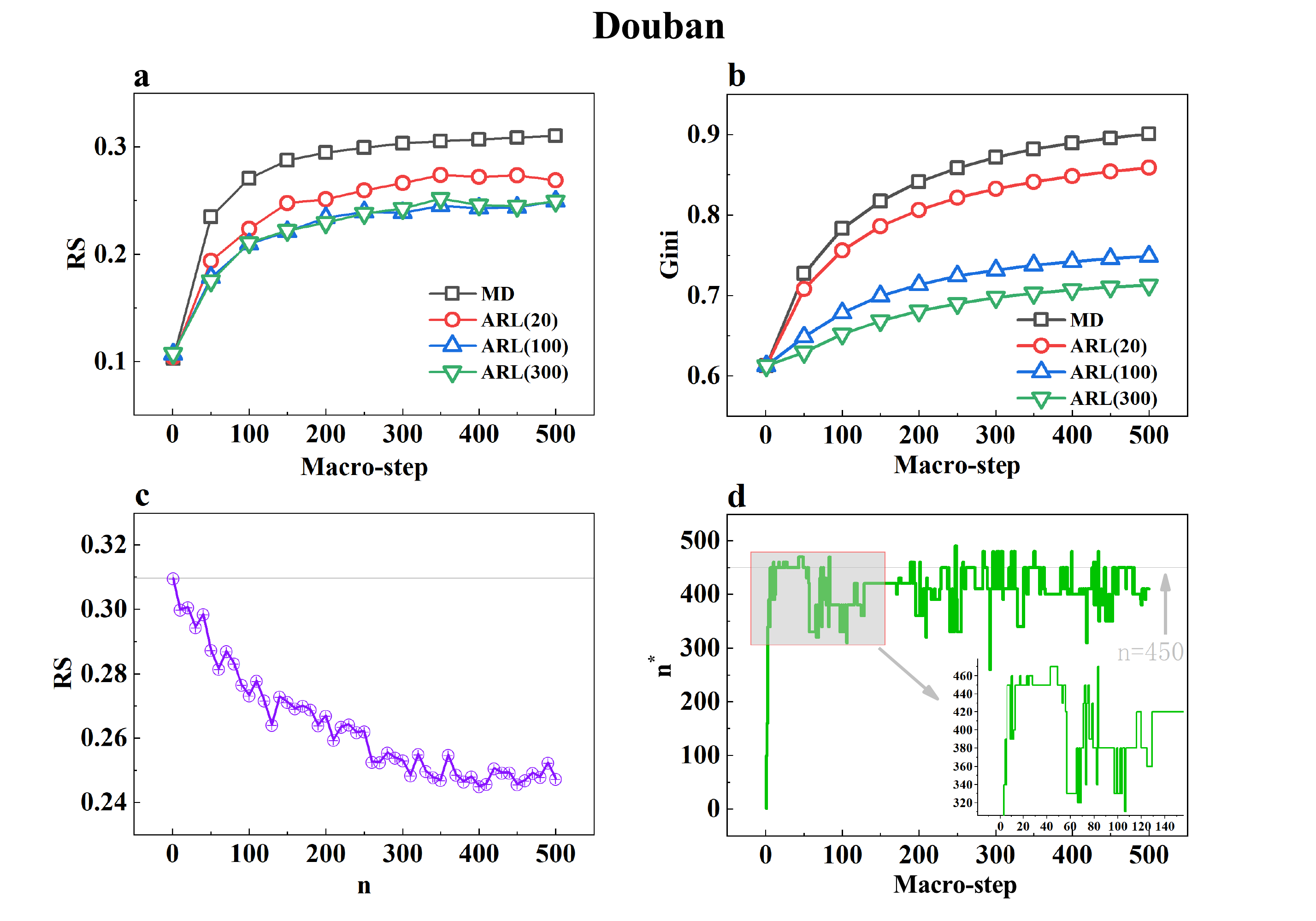}
		\caption{Douban. \textbf{\emph{(a)(b)(c)(d)}} The description is the same as the caption of Fig.6 .}
		\label{Fig.S5}
\end{figure}
	
The result obtained from the Douban data is shown in the \autoref{Fig.S5}. The long-term accuracy and Gini index can be improved greatly by ARL in \autoref{Fig.S5} (a)(b). In addition, those parameters to enhance long-term accuracy appear in larger interval in \autoref{Fig.S5} (c). The similar result can be observed in more dense dataset, such as Epinions, Netflix and Movilens (Fig.4 in main text). Meanwhile, the accuracy of most parameters keep the small difference with the optimal parameter in the long-term recommendation. In the \autoref{Fig.S5} (d), the most optimal parameters under different macro step fluctuates between 300 and 500. Actually, the difference of long-term recommendation accuracy is very small (see in \autoref{Fig.S5} (c)), which confirms again that ARL method is very robust.
	
\section*{The effect of  data set on $n^{*}$}
We conducted experiments on different datasets and locate the optimal value of parameter ($n^*$) according the current resolution (a step of 10). We use the sparsity to quantify the dataset. Sparsity is calculated by the formula \autoref{E1}: 
\begin{equation}
	\label{E1}
	Sparsity = \frac{N_{link}}{N_{item}*N_{user}}                  
\end{equation}

The detailed statistical characteristics is shown in \autoref{S1}. Then we analyze the spearman correlation between $n^*$ and sparsity (\autoref{S2}), finding there is a strong correlation between the optimal parameters and the sparsity of the network. This indicate that the value of optimal parameter $n^*$ is relate to sparsity of datasets. This is consistent with our expectation because the coverage of initial resource is limited in sparse dataset. In such condition, we need to introduce similar items into evolving network to keep user's preference instead of more diverse items. However, diverse items need to be added into the evolving network on the denser dataset, which further distribute the initial resource to cold items.
	
\begin{table}[!ht]
	\setlength{\abovecaptionskip}{3pt}
	\caption{The statistical characteristics of experimental dataset}
	\label{S1}
	\centering
	\small
\begin{tabular}{clclc}
		\hline
		\multicolumn{2}{c}{\multirow{2}{*}{\textbf{Datasets}}} & \multicolumn{2}{c}{\multirow{2}{*}{\textbf{$n^*$}}} & \multirow{2}{*}{\textbf{Sparsity}} \\
		\multicolumn{2}{c}{}                                   & \multicolumn{2}{c}{}                            &                                    \\ \hline
		\multicolumn{2}{c}{Delicious}                          & \multicolumn{2}{c}{20}                          & $0.20\times10^{-2}$                               \\
		\multicolumn{2}{c}{Amazon}                             & \multicolumn{2}{c}{30}                          & $0.22\times10^{-2}$                               \\
		\multicolumn{2}{c}{Stack Overflow}                     & \multicolumn{2}{c}{150}                         & $0.33\times10^{-2}$                               \\
		\multicolumn{2}{c}{Epinions}                           & \multicolumn{2}{c}{470}                         & $1.36\times10^{-2}$                               \\
		\multicolumn{2}{c}{Netflix}                            & \multicolumn{2}{c}{270}                         & $1.79\times10^{-2}$                               \\
		\multicolumn{2}{c}{Douban}                             & \multicolumn{2}{c}{400}                         & $2.63\times10^{-2}$                               \\
		\multicolumn{2}{c}{Movielens}                          & \multicolumn{2}{c}{320}                         & $3.49\times10^{-2}$                               \\ \hline
	\end{tabular}
\end{table}
	
\begin{table}[!ht]
	\setlength{\abovecaptionskip}{3pt}
	\caption{The Spearman correlation between $n^*$ and sparsity}
	\label{S2}
	\centering
	\small
	\begin{tabular}{|l|l|c|c|}
		\hline
		\multicolumn{2}{|l|}{\multirow{2}{*}{\textbf{Spearman correlation}}} & \multirow{2}{*}{\textbf{$n^*$}} & \multirow{2}{*}{\textbf{Sparsity}} \\
		\multicolumn{2}{|l|}{}                                               &                              &                                    \\ \hline
		\multicolumn{2}{|c|}{\textbf{$n^*$}}                                    & 1                            & 0.75                               \\ \hline
		\multicolumn{2}{|c|}{\textbf{Sparsity}}                              & 0.75                         & 1                                  \\ \hline
		\end{tabular}
	\end{table}
	
	
\end{document}